\documentclass{aa}
\usepackage{graphicx}
\usepackage{txfonts}
\usepackage{longtable}
\usepackage{lscape}
\usepackage{natbib}
\usepackage{multirow}

\def\kms{km.s$^{-1}$}         %m.s -1
\def\ms{\hbox{m.s$^{-1}$}}         %m.s -1
\def\kms{\hbox{km.s$^{-1}$}}       %km.s -1
\def\degr{\hbox{$^\circ$}}
\def\teff{T$_{\rm eff}$}
\def\logg{log~{\it g}}
\def\met{[Fe/H]}

\def\logrhk{log~R$'_{\rm HK}$}

\begin{document}

   \title{K2-110 b -- a massive mini-Neptune exoplanet}

   %\subtitle{I. Overviewing the $\kappa$-mechanism}
   \author{H.~P.~Osborn\inst{\ref{warwick}}\thanks{email: h.p.osborn@warwick.ac.uk} %first author
          \and A.~Santerne\inst{\ref{IA}}\fnmsep\inst{\ref{LAM}} %HARPS PI
          \and S.~C.~C.~Barros\inst{\ref{IA}} % HARPS CoI
          \and N.~C.~Santos\inst{\ref{IA}}\fnmsep\inst{\ref{UPorto}} % HARPS CoI
          \and X. Dumusque\inst{\ref{geneva}}
              \and L. Malavolta\fnmsep\inst{\ref{padova}}\inst{\ref{padova2}} % HARPS-N CoI 
          \and D.~J.~Armstrong\inst{\ref{warwick}}\fnmsep\inst{\ref{Belfast}} % HARPS CoI
          \and S.~Hojjatpanah\inst{\ref{IA}}\fnmsep\inst{\ref{zanjan}} % work on the HARPS data
           \and O.~Demangeon\inst{\ref{LAM}} % HARPS CoI
           \and V.~Adibekyan\inst{\ref{IA}} % spectral analysis
          \and J.-M.~Almenara\inst{\ref{Grenoble}}\fnmsep\inst{\ref{ipag}} % HARPS CoI
          \and D.~Barrado\inst{\ref{CAB}} % HARPS CoI
          \and D.~Bayliss\inst{\ref{geneva}} % CORALIE screening
          \and I.~Boisse\inst{\ref{LAM}} % HARPS CoI
          \and F.~Bouchy\inst{\ref{geneva}}\fnmsep\inst{\ref{LAM}} % HARPS coordinator
          \and D.~J.~A.~Brown\inst{\ref{warwick}} % HARPS CoI
          \and A. C. Cameron\inst{\ref{StAnd}} % HARPS-N CoI
          \and D. Charbonneau\inst{\ref{CFA}} % HARPS-N CoI
          \and M.~Deleuil\inst{\ref{LAM}} % HARPS CoI
          \and E.~Delgado~Mena\inst{\ref{IA}} % spectral analysis
          \and R.F. ~D\'iaz\inst{\ref{geneva}}\fnmsep\inst{\ref{UBA}}\fnmsep\inst{\ref{IAFE}} % Pastis
          \and G.~H\'ebrard\inst{\ref{IAP}}\fnmsep\inst{\ref{OHP}} % HARPS CoI
          \and J.~Kirk\inst{\ref{warwick}} % WRK lightcurve eyeballer
          \and G.~W.~King\inst{\ref{warwick}} % WRK lightcurve eyeballer (actually spotted this planet)
          \and K.~W.~F.~Lam\inst{\ref{warwick}} % WRK lightcurve eyeballer
          \and D.~Latham\inst{\ref{CFA}} % HARPS-N CoI
          \and J.~Lillo-Box\inst{\ref{ESO}} % HARPS CoI
          \and T.~M.~Louden\inst{\ref{warwick}} % WRK eyeballing team
          \and C.~Lovis\inst{\ref{geneva}} % HARPS coordinator
          \and M.~Marmier\inst{\ref{geneva}} % CORALIE screening
          \and J.~McCormac\inst{\ref{warwick}} % WRK lightcurve eyeballer
          \and E. Molinari\inst{\ref{TNG}}\fnmsep\inst{\ref{IASF}} % HARPS-N CoI 
          \and F.~Pepe\inst{\ref{geneva}} % HARPS-N CoI 
          \and D.~Pollacco\inst{\ref{warwick}} % HARPS CoI 
          \and S.~G.~Sousa\inst{\ref{IA}} % spectral analysis
          \and S.~Udry\inst{\ref{geneva}} % CORALIE screening
          \and S.~R.~Walker\inst{\ref{warwick}} % WRK eyeballing team
          }

   \institute{Department of Physics, University of Warwick, Gibbet Hill Road, Coventry, CV4 7AL, UK\label{warwick}
              \and
              Instituto de Astrof\'isica e Ci\^{e}ncias do Espa\c co, Universidade do Porto, CAUP, Rua das Estrelas, 4150-762 Porto, Portugal\label{IA}
              \and
              Aix Marseille Universite, CNRS, Laboratoire d'Astrophysique de Marseille UMR 7326, 13388, Marseille, France\label{LAM}
              \and
               Departamento de F\'isica e Astronomia, Universidade do Porto, Rua Campo Alegre, 4169-007 Porto, Portugal\label{UPorto}
              \and
              Observatoire Astronomique de l'Universite de Geneve, 51 Chemin des Maillettes, 1290 Versoix, Switzerland\label{geneva}
              \and
              Dipartimento di Fisica e Astronomia "Galileo Galilei" , Universita’ di Padova, Vicolo dell’Osservatorio 3, 35122 Padova, Italy \label{padova}
              \and
              INAF - Osservatorio Astronomico di Padova, Vicolo dell'Osservatorio 5, 35122 Padova, Italy\label{padova2}
              \and
              ARC, School of Mathematics \& Physics, Queen's University Belfast, University Road, Belfast BT7 1NN, UK\label{Belfast}
              \and
              Department of Physics, University of Zanjan, University Blvd, 45371-38791, Zanjan, Iran\label{zanjan}
              \and
              Universite Grenoble Alpes, IPAG, 38000 Grenoble, France\label{Grenoble}
              \and
              CNRS, IPAG, 38000 Grenoble, France\label{ipag}
              %\and
               Depto. de Astrof\'isica, Centro de Astrobiolog\'ia (CSIC-INTA), ESAC campus 28692 Villanueva de la Ca\~nada (Madrid), Spain\label{CAB}
               %Depto. Astrof\'{\i}sica, Centro de Astrobiolog\'{\i}a (INTA-CSIC),  ESAC campus, Camino Bajo del Castillo s/n, E-28692 Villanueva de la Ca\~nada, Spain\label{CAB}
%              Departamento de Astrofsica, Centro de Astrobiologa (CSICINTA), ESAC campus 28692 Villanueva de la Caada (Madrid), Spain\label{CAB}
              \and
              School of Physics \& Astronomy, University of St. Andrews, North Haugh, St. Andrews Fife, KY16 9SS, UK \label{StAnd}
              \and
              Harvard-Smithsonian Center for Astrophysics, 60 Garden Street, Cambridge, Massachusetts 02138, USA \label{CFA}
              \and
              Universidad de Buenos Aires, Facultad de Ciencias Exactas y Naturales. Buenos Aires, Argentina.\label{UBA}
              \and
              CONICET - Universidad de Buenos Aires. Instituto de Astronom\'ia y F\'isica del Espacio (IAFE). Buenos Aires, Argentina. \label{IAFE}
              \and
              Institut d'Astrophysique de Paris, UMR7095 CNRS, Universite Pierre \& Marie Curie, 98bis boulevard Arago, 75014 Paris, France\label{IAP}
              \and
              Observatoire de Haute-Provence, Universite d\'Aix-Marseille \& CNRS, 04870 Saint Michel l'Observatoire, France\label{OHP}
              \and
              European Southern Observatory (ESO), Alonso de Cordova 3107, Vitacura, Casilla 19001, Santiago de Chile, Chile\label{ESO}
              \and 
              INAF - Fundación Galileo Galilei, Rambla José Ana Fernandez Pérez 7, 38712 Berña Baja, Spain \label{TNG}
              \and
              INAF - IASF Milano, via Bassini 15, 20133, Milano, Italy \label{IASF}
              }

   \date{Submitted May 13, 2016}

% \abstract{}{}{}{}{} 
% 5 {} token are mandatory
 
  \abstract{We report the discovery of the exoplanet K2-110 b (previously EPIC212521166b) from K2 photometry orbiting in a 13.8637d period around an old, metal-poor K3 dwarf star.
  With a V-band magnitude of 11.9, K2-110 is particularly amenable to RV follow-up.
  A joint analysis of K2 photometry and high-precision RVs from 28 HARPS and HARPS-N spectra reveal it to have a radius of 2.6$\pm 0.1$~R$_{\oplus}$ and a mass of 16.7$\pm 3.2$~M$_{\oplus}$, hence a density of $5.2\pm1.2$~g.cm$^{-3}$, making it one of the most massive planets yet to be found with a sub-Neptune radius.
  When accounting for compression, the resulting Earth-like density is best fitted by a $0.2$~M$_{\oplus}$ hydrogen atmosphere over an $16.5$~M$_{\oplus}$ Earth-like interior, although the planet could also have significant water content.
  At 0.1~AU, even taking into account the old stellar age of $8 \pm 3$~Gyr, the planet is unlikely to have been significantly affected by EUV evaporation. However the planet likely disc-migrated to its current position making the lack of a thick H$_2$ atmosphere puzzling.
  This analysis has made K2-110 b one of the best-characterised mini-Neptunes with density constrained to less than 30\%.
}
   %{context}
   %{Aims to go in here}
   %{methods to go in here}
   %{results in here}
%   {conclusions in here}

   
   \titlerunning{Detection of K2-110 b}
   %\titlerunning{EPIC212521166b - Detection}
   \authorrunning{HP Osborn et al}
   \maketitle
%
%________________________________________________________________
%@arxiver{TransitBestFits_PastisPosts_RR,EPIC-1166_RVphase_RR,MR_Diagram_RR.pdf} 

\section{Introduction}
Since 2014, \textit{Kepler}'s extended \textit{K2} mission \citep{howell2014k2} has observed eleven 80-day fields, giving precise photometry for more than 200,000 stars.
So far, it has produced around 500 planet candidates \citep{foreman2015systematic, vanderburg2015,barros2016new, pope2016transiting} and nearly 150 confirmed or validated planets \citep{montet2015stellar,sinukoff2015ten,armstrong2015one,barros2015photodynamical,crossfield2016197}. \textit{K2} has also significantly expanded the population of small planets transiting bright stars, with the number of \textit{K2} planets around stars with \textit{Kepler} magnitude $11-12.5$ already exceeding the initial four-year \textit{Kepler} mission \citep{crossfield2016197}.
\begin{figure*}
\makebox[\textwidth][c]{
 \includegraphics[width=\paperwidth]{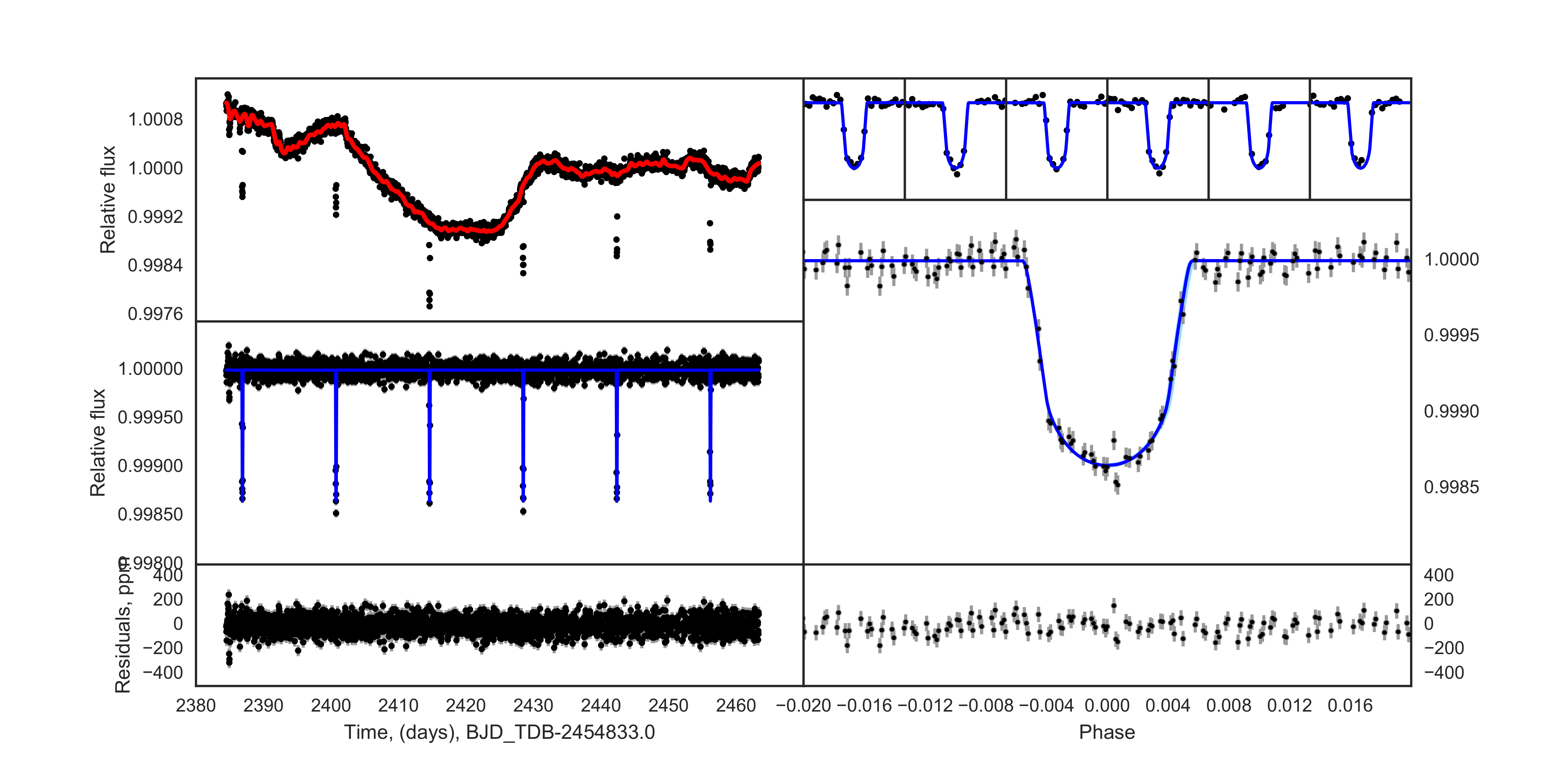}
}
\caption{Transit lightcurve and model best-fit.
Detrended \textit{K2} Lightcurve (upper left); Lightcurve smoothed with Gaussian Processes (centre left); best-fit transit model lightcurve residuals (lower left); All six transits (upper right); Phase-folded unbinned lightcurve centred on transit with best-fit model in blue \& best-fit region in light blue (centre right); Phase-folded model residuals (lower right).}
% * <alexandre.santerne@gmail.com> 2016-04-26T18:00:49.991Z:
%
% You should put the phot residuals in ppm to make the plot easier to read
%
% ^.
\label{FullTransitFits0}
\end{figure*}

Large super-Earths and mini-Neptunes are an interesting and diverse population to study.
Lying in the transition regime between terrestrial planets and gas giants, they can provide direct constraints on giant planet formation theory \citep[e.g. ][]{hansen2012migration}.
They range from low-density neptunes \citep[e.g. the Kepler-11 system; $0.6 - 1.7$~g.cm$^{-3}$ ][]{lissauer2013all}, to super-Earths with large rocky interiors \citep[e.g. Kepler-10 c; $7\pm1$~g.cm$^{-3}$][]{Kep10c}, to controversial claims of ultra-dense super-Earths that could be evaporated gas giant cores \citep[e.g. K2-38 b; $17.5\pm7$~g.cm$^{-3}$][]{sinukoff2016eleven}.
With magnitudes brighter than V=12, such planets are also suitable targets for ground-based follow-up targeting radial velocities (RVs), the Rossiter-McLaughlin effect and atmospheric spectroscopy.
We present the detection of a sub-Neptune radius planet around the K field dwarf K2-110 from \textit{K2} photometry and the confirmation and mass measurement of this planet with HARPS and HARPS-N RVs.

%Transiting exoplanets allow unique insight into the radius (from photometry) and  and  have been observed to have a distinct boundary between rocky composition and gaseous
%Other transiting warm Neptunes to mention?
%Kepler-131b
%Kepler-20c
%CoRoT-22b (11.93 mag, G0, supEarth)
%HD97658b
%K2-19c

%__________________________________________________________________

\section{Observations, data reduction and analysis}
\subsection{K2 Photometry}
   
EPIC212521166 was observed during Campaign 6 of the \textit{K2} mission. We downloaded the pixel data from the Mikulski Archive for Space Telescopes (MAST)\footnotemark \footnotetext{$http://archive.stsci.edu/kepler/data\_search/search.php$}
and used a modified version of the CoRoT imagette pipeline to extract the light curve. 
Based on signal-to-noise of each pixel we computed an optimal aperture of 25 pixels. The background was estimated using the $3\sigma$ clipped median of all the pixels in the image and subtracted. We also calculated the centroid using the Modified Moment Method by \citet{Stone1989}.
In order to correct for flux variations due to the star's apparent motion on the charge-coupled device (CCD) we used a self-flat-fielding procedure similar to \citet{vanderburg2014technique}. 
This assumes the movement of the satellite was mainly in one direction, as described in \citet{barros2016new}. 
The lightcurve was flattened with an exponential square Gaussian process trained on out-of-transit data. Transit fitting was performed jointly with RVs and is described in section 2.4.
The final light curve of K2-110 has mean out-of-transit RMS of $134\,$ parts per million (ppm) per 30~min cadence. 

%______________________________________________ 
\subsection{Detection}

EPIC212521166 was indentified as a strong planet candidate during two independent searches for transits in the lightcurve. 
Both \citet{barros2016new} using the CoRoT alarm pipeline and the detrending method defined above, and a manual search of bright Campaign 6 lightcurves detrended using the technique of \cite{armstrong2015k2} detected K2-110 b.
It was also independently identified by \citet{aigrain_2016} \& \citet{pope2016transiting} although no detailed analysis was performed.

%______________________________________________ 
\subsection{Radial velocity follow-up}

We performed RV follow-up of the target star.
A single exposure with the CORALIE spectrograph \citep{queloz2000coralie} mounted on the EULER telescope at ESO La Silla observatory, Chile confirmed that the target was suitable for precise radial velocities. % A second exposure would have ruled out spectroscopic binaries but given the 14-d period of the orbit, this second epoch was not possible between the detection of the candidate in the K2 data and the telescope scheduling (you can remove this last sentence if you dislike it)}. 

The target was then observed with the HARPS spectrograph \citep{mayor2003setting}, mounted on the 3.6m telescope at ESO La Silla observatory\footnote{ESO programme ID: 096.C-0657}. 
Seventeen exposures of 3600~s in the obj\_AB mode were secured on 13 nights from 2016-03-03 to 2016-08-10, with S/N per pixel at 5500~\r{A} from 25 to 57 leading to photon noise uncertainty ($\sigma_{\rm RV}$) in the range 1.2 -- 3.0~\ms. 

Eleven further spectra were also taken with the HARPS-N spectrograph \citep{cosentino2012harps}, mounted on the 3.58m Telescopio Nazionale Galileo  at ING La Palma observatory.
These had a median exposure time of 1800s each and were secured on 11 nights from 2016-05-12 to 2016-06-25, with S/N per pixel  at order $n=50$ (5500~\r{A}) from 15 to 37. The HARPS-N exposures suffered from contamination of moonlight, which were corrected by subtracting the CCF of the spectrum of the sky (gathered with fiber B) from the CCF of the target star (gathered with fiber A), after recomputing the former with the same flux correction coefficients of the latter (see \citet{2017arXiv170306885M} for details). Although contamination corrections were applied to all HARPS-N measurements, only six of these had absolute RV corrections ($|\Delta$RV$|$) greater than the photon noise ($\sigma_{\rm RV}$), with four being corrected by more than $2~\sigma_{RV}$ (see Table 4).

We computed the RVs from each high-resolution spectrum using weighted cross-correlation with a K5 template \citep{baranne1996elodie,pepe2002harps} as implemented by the HARPS and HARPS-N pipelines. 
RV uncertainties were determined as described in \citet{bouchy2001fundamental}. They range from 1.1 to 8.8~m.s$^{-1}$. The pipeline also computed the averaged line profile full-width half maximum (FWHM), bisector span and the \logrhk. All these data are reported with uncertainties in Appendix Table \ref{RVtab}.

   \begin{figure}
   \centering
   \includegraphics[width=\hsize]{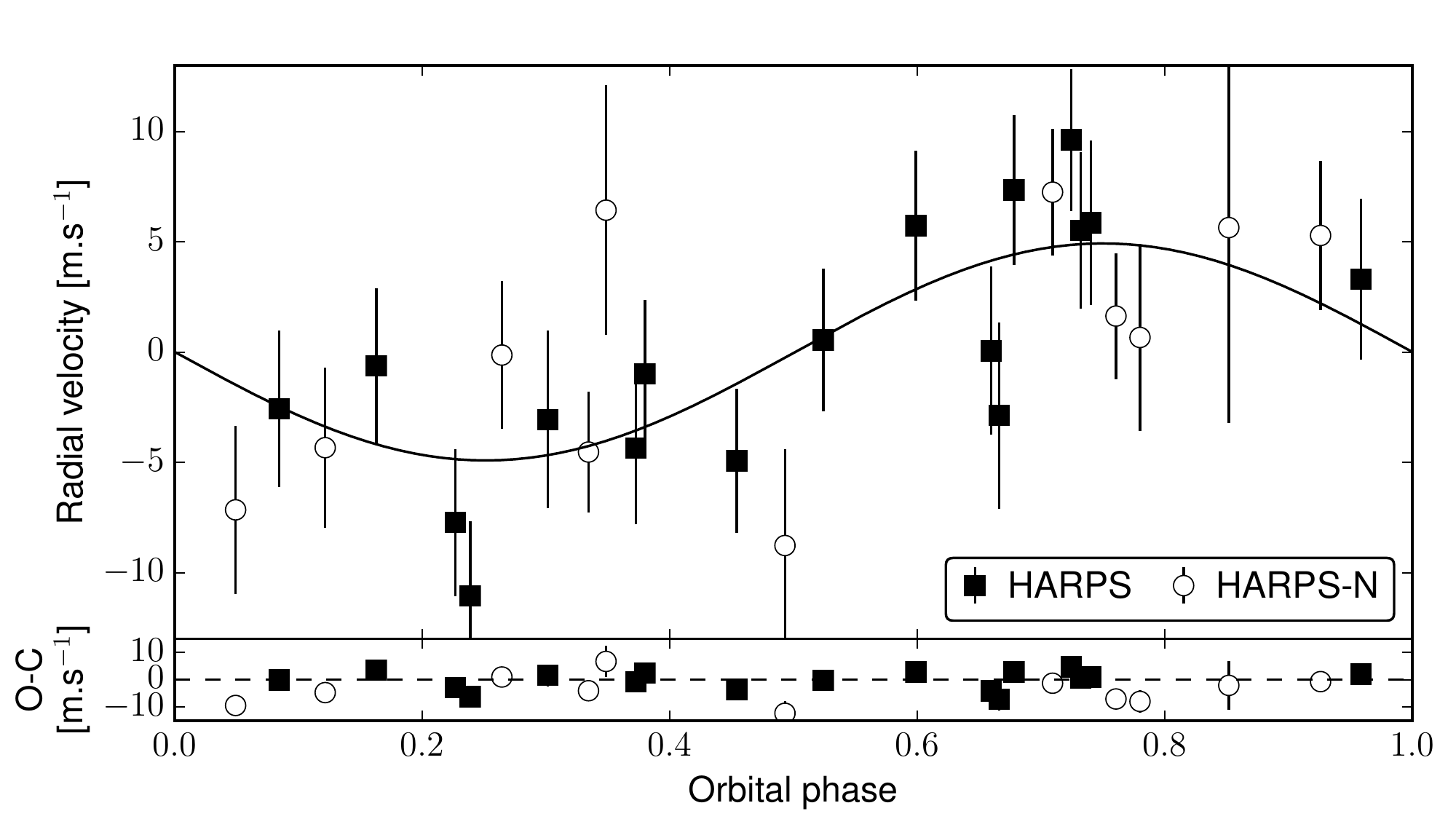}
      \caption{RVs taken by HARPS and HARPS-N, phased to the 13.86d period as determined from the \textit{K2} lightcurve. The best model is also displayed together with the residuals.% which present a RMS of 2.0ms$^{-1}$, which is compatible with the mean photon noise of 1.8ms$^{-1}$.
              }
         \label{HarpsRVs}
   \end{figure}

% \begin{figure}
% \centering      
%   \resizebox{\textwidth}{!}{\includegraphics[\textwidth][bb= 0 600 1000 0,clip]{TransitBestFits_PastisPosts.pdf}}
% %  \includegraphics[bb=10 20 100 300,clip]{TransitBestFits_PastisPosts.pdf}
%     \caption{Transit lightcurve and model best-fit.}
%     \label{FullTransitFits}
% \end{figure}

%__________________________________________________ 
\subsection{Host star parameters}
%Removed old stellar params. Some dont agree...
% Teff    & 4960  &   $\pm 60$\\
% log(g)  & 4.58  &   $\pm 0.13$\\
% Fe\/H  & -0.34  & $\pm 0.03$\\
% V\_t     & 0.57   &   $\pm 0.17$\\
% Mass  &  0.75 & $\pm 0.02$\\
% Radius &  0.74 & $\pm 0.12$\\

%From Nuno:

%Santos-2013,Santos-2015,Sneden-1973,Kurucz-1993,Sousa-2011,Tsantaki-2013,Adibekyan-2015,

Stellar atmospheric parameters and [Fe/H] were derived in LTE using a recent version of the MOOG code \citep{Sneden-1973} and a set of plane-parallel ATLAS9 model atmospheres \citep{Kurucz-1993}, as described in \cite{sousa2011spectroscopic}. 
The full spectroscopic analysis is based on the equivalent widths (EWs) of 103 \ion{Fe}{i} and 15 \ion{Fe}{ii} weak lines by imposing ionization and excitation equilibrium. The line-list used was taken from \cite{Tsantaki-2013}, and is adapted for stars with $T_{\text{eff}}<5200$~K. 
The stellar parameters derived using this methodology were shown to be in line with the ones derived using other methods, for example, interferometry and Infra-Red Flux Method (IRFM) \citep[see][for details]{Santos-2013}.
Our derived values of Teff ($4960\pm60$~K), log(g) ($4.58\pm0.13$) and [Fe/H] ($-0.34\pm0.03$) are used as priors in \texttt{PASTIS} (see Section 2.5 and Table 5) and therefore are re-derived in a self-consistent manner with all observed stellar information. We also find a \logrhk of $-4.983\pm0.002$.
% log(g)  = 4.58      +- 0.13
%Results are shown in Table \ref{physics} and 
Chemical abundances for different species derived using the methodology described in, for example \cite{adib2016} and \cite{Santos-2015} are shown in in Appendix Table \ref{abunds}.

%Similar properties compared with the other similar planets that are well characterised: HD97658 b (Vmag 7, 7Me/2.3Re) and HIP 116454b/K2-2b (Vmag 10.2 11Me/2.5Re): The three planets transit metal-poor early-K dwarfs. They have a similar composition (half rock, half water) -- see attached. \texbtf{move to the discussion ?}

%__________________________________________________
%\subsection{Lucky Imaging}

%__________________________________________________
\subsection{PASTIS analysis}

% from Alex
We jointly analysed the HARPS/HARPS-N RVs, \textit{K2} photometric light curve within 3h of the transit midtime and the spectral energy distribution (SED) as observed by the APASS, 2-MASS and WISE surveys \citep{munari2014apass,cutri2014vizier} using the \texttt{PASTIS} software \citep{diaz2014pastis,santerne2015pastis}.
It models the light curve using the \texttt{jktebop} package \citep{southworth2008homogeneous} assuming an oversampling factor \citep{kipping2010binning} of 30 to account for the long integration time of the K2 data. 
The SED was modelled using the BT-SETTL library of stellar atmosphere \citep{allard2012models}. The RVs were modelled with a Keplerian orbit. 
A Markov Chain Monte Carlo (MCMC) method was used to analyse the data. 
The results from the spectral analysis described in Section 2.4 were used as priors for the host star. 
The spectroscopic parameters were converted into fundamental stellar parameters in the MCMC using the Darthmouth evolution tracks \citep{dotter2008dartmouth}.
For detail on the priors used see Table 5.%, \citet{diaz2014pastis}, and \citet{santerne2015pastis}.

\begin{table}
\caption{Stellar Information for K2-110. Magnitudes from ExoFOP-K2.
}
\label{stellarinfo}
\centering
\begin{tabular}{l c}     % 3 columns 
\hline
\hline
Parameter & Value and uncertainty\\
\hline
\multicolumn{2}{l}{\it Stellar information}\\
 &\\
EPIC & $ 212521166 $\\
R.A. &  13h 49m 23.890s \\
Dec &  -12d  17m  04.16s \\
2MASS ID & 2MASS J13492388-1217042 \\
$\mu_{\text{R.A.}}$ & $42.6 \pm 1.1$ mas/yr \\
$\mu_{\text{Dec}}$ & $-101.2 \pm 1.4$ mas/yr \\
 &\\
\hline
\multicolumn{2}{l}{\it Photometric magnitudes}\\
 &\\
B & $ 12.834 \pm 0.05 $\\
V & $ 11.91 \pm 0.07 $\\
Kep & $11.59$ \\
J &$ 10.184 \pm 0.022$\\
H & $9.641 \pm 0.023 $ \\
K & $ 9.607 \pm 0.024$\\
WISE $3.4\mu$m & $9.521 \pm 0.024$ \\
WISE $4.6\mu$m & $9.577 \pm 0.020$ \\
WISE $12\mu$m & $9.479 \pm 0.038$ \\
WISE $22\mu$m & $8.695$ \\
 & \\
\hline
\hline
\end{tabular}
\end{table}

%Alex: new table
\begin{table}%[h]
\caption{Physical parameters of the K2-110 system}
\begin{center}
\setlength{\tabcolsep}{0.9mm}
\begin{tabular}{lc}
\hline
\hline
Parameter & Value and uncertainty\\
\hline
\multicolumn{2}{l}{\it Stellar parameters}\\
 &\\
 Stellar mass $M_{\star}$ [M$_\odot$] & 0.738 $\pm$ 0.018\\
 Stellar radius $R_{\star}$ [R$_\odot$] & 0.713 $\pm$ 0.020\\
 Stellar age $\tau$ [Gyr] & 8 $\pm$ 3\\
 Distance $d$ [pc] & 118.0 $\pm$ 3.6\\
 Reddening E(B-V) [mag] & 0.019$^{_{+0.019}}_{^{-0.013}}$\\
 Systemic RV $\upsilon_{0}$ [km.s$^{-1}$] & -21.6331 $\pm 9\times10^{-4}$\\
 Effective temperature T$\mathrm{eff}$ [K] & 5010 $\pm$ 50\\
 Surface gravity log$g$ [g.cm$^{-2}$] & 4.60 $\pm$ 0.03\\
 Iron abundance [Fe/H] [dex] & -0.34 $\pm$ 0.03\\
% Rotational velocity $\upsilon\sin i$ [km.s$^{-1}$] & 3.7 $\pm$ 0.5\\
% Rotation period P$_{\rm rot}$ [d] & 10.79 $\pm$ 0.02\\
% Stellar luminosity $L/L_{\astrosun}$ & 0.56 $\pm$ 0.02\\
 Spectral type & K3V\\
 &\\
 \hline
\multicolumn{2}{l}{\it Orbital parameters}\\
 &\\
Period $P$ [d] & 13.86375 $\pm$ 2.6$\times10^{-4}$\\
Transit epoch T$_{0}$ [BJD$_\mathrm{TDB}$] & 2457275.32992 $\pm$ 6.1$\times10^{-4}$\\
Orbital eccentricity $e$ & 0.079$\pm0.07$\\
Argument of periastron $\omega$ [\degr] & 90$^{_{+180}}_{^{-64}}$\\
Inclination $i$ [\degr] & 89.35 $^{_{+0.41}}_{^{-0.24}}$\\
Semi-major axis $a$ [AU] & 0.1021 $\pm$ 8$\times10^{-4}$\\
 &\\
\hline
\multicolumn{2}{l}{\it Transit \& radial velocity parameters}\\
 &\\
 System scale $a/R_{\star}$ & 30.8 $\pm$ 1.0\\
 Impact parameter $b$ & 0.34 $^{_{+0.14}}_{^{-0.22}}$\\
 Transit duration T$_{14}$ [h] & 3.22 $\pm$ 0.03\\
 Planet-to-star radius ratio $k_{r}$ & 0.0333 $\pm$ 6.6$\times10^{-4}$\\
 Limb darkening $u_a$ & $0.5322 \pm 1.2\times10^{-2}$ \\
 Limb darkening $u_b$ & $0.1787 \pm 8\times10^{-3}$ \\
 RV amplitude $K$ [m.s$^{-1}$] & $5.5 \pm 1.1$\\
 %RV -- S$_{MW}$ correlation & -0.153 $\pm$ 0.057\\
 %S$_{MW}$ offset & -0.1837 $\pm$ 0.020\\
 HARPS-N RV jitter [m.s$^{-1}$] & 3 $\pm$ 2 \\ %(included in the error of the plot)
 HARPS RV jitter [m.s$^{-1}$] & 3.1 $\pm$ 1 \\% (included in the error of the plot)
 %HARPS-N RVoffset relative to HARPS: 0.0042 +/- 1.9e-03 km/s
 Instrument offset  [m.s$^{-1}$] & 4.2 $\pm$ 1.8\\
 K2 contamination  [flux, ppt] & $3.4^{+4}_{-2}$ \\
 K2 jitter  [flux, ppm] & $40 \pm 4.6$ \\
 SED jitter [mag]& 0.02 $\pm$ 0.02 \\
 &\\
\hline
\multicolumn{2}{l}{\it Planet parameters}\\
 &\\
 Planet mass $M_{p}$ [M$_\oplus$] & 16.7 $\pm$ 3.2\\
 Planet radius $R_{p}$ [R$_\oplus$] & 2.592 $\pm$ 0.098\\
 Planet density $\rho_{p}$ [g.cm$^{-3}$] & 5.2 $\pm$ 1.2\\
 Equilibrium temperature $T_{\rm eq}$ [K] & 640 $\pm$ 20 \\
 &\\
 \hline
 \hline
\end{tabular}
\end{center}
\label{physics}
\tablefoot{All the uncertainties provided here are only the statistical ones. Errors on the models are not considered, as they are unknown. Stellar parameters are derived from the combined analysis of the data and not from the spectral analysis. We assumed R$_\odot$=695 508km, M$_\odot$=1.98842$\times10^{30}$kg, R$_\oplus$=6 378 137m, M$_\oplus$=5.9736$\times10^{24}$kg, and 1AU=149 597 870.7km.}
\end{table}%

We used uninformative priors for most of the parameters. Exception include the stellar atmospheric parameters, for which we used the inputs of Section 2.4; the orbital eccentricity, for which we choose a Beta distribution \citep{kipping2013parametrizing}; and the orbital ephemeris, for which we choose uniform distributions centred on the values found by the detection pipeline.% and assuming a width of 0.01 d and 0.1 d for the period and transit epoch, respectively. 

We ran 20 independent MCMCs of $3\times10^5$ iteration randomly started from the joint prior distribution. We then removed the burn-in phase before merging the converged chains (see \citet{diaz2014pastis}). The residuals of the RV have a RMS at the level of 3.0~m.s$^{-1}$, which is approximately twice the median photon noise. %The residuals exhibit a clear correlation with the Mount Wilson S index with a Pearson correlation coefficient of $\rho_{S_{MW}} = 0.74 \pm 0.07$. We find however no correlation between the radial velocity residuals and the FWHM or the bisector with Pearson test values of $\rho_{FWHM} = -0.24 \pm 0.07$ and $\rho_{BIS} = 0.06 \pm 0.04$.
The fit was also run with uniform priors on both eccentricity and limb darkening parameters to test if these may be influencing the resulting planet parameters. However, all outputs were consistent within 1-sigma to those obtained during the initial \texttt{PASTIS} fit.
Complete stellar and planetary outputs of \texttt{PASTIS} are reported in Table \ref{physics}.

%__________________________________________________

\section{Discussion}
\subsection{Validity}
The presence of a RV signal in-phase with (and at a similar amplitude to) that expected from the transit detection is extremely strong evidence for a planet.
However, we also performed additional tests to ensure that the signal was not due to, for example, a blended eclipsing binary.

Using the cross-correlation function with which we computed RVs, we can exclude to 3-sigma all secondaries with $\Delta$mag$ < 6.5$, assuming the companion is spectrally resolved (ie $v_0>2.7$~km.s$^{-1}$), and has a similar Teff, metallicity, and rotation \citep{santerne2015pastis}.
%We can also rule out close companions within 3arcsecs from archival survey data \citep{2mass,usno}.
Although \textit{K2}'s pixel drift is significant, we can constrain in-transit centroid shifts that result from a signal from a background star to 0.86" and 0.28" (2-sigma limits) in the x \& y directions.
%ALEX's sentence on BIS, etc, to go here.
The BIS and FWHM do not exhibit significant variation with RMS at the level of 6m.s$^{-1}$ and 5m.s$^{-1}$, respectively. No significant correlation is found between these spectroscopic diagnoses and the radial velocity with a Pearson test of 0.01 $\pm$ 0.02, and 0.13 $\pm$ 0.02, respectively. This strongly supports the planetary nature of the detection \citep{santerne2015pastis}.
Small Neptunes also have the lowest false positive rates (6.7\% in \textit{Kepler},  \citealt{fressin2013false}). 
Together all these are extremely good evidence that the signal is planetary rather than from a false positive, and enables us to designate this planet as confirmed.

\subsection{Age and rotation}
We derived
a stellar age of $8 \pm 5$ Gyr from the joint orbital analysis of RVs, photometry, and stellar evolution tracks, with posterior samples cut such that age $<13.5$~Gyr, .
No clear rotational signal is detected in the lightcurve, although variation on the order of weeks is seen which could be suggestive of slow starspot rotation (and therefore an old gyrochronological age).
%We also detected a $50\pm5$d sinusoidal variation in the lightcurve with a Lomb-Scargle search which, if attributed to starspot rotation features, would suggest an age of $11\pm 5$Gyr \citep{angus2015calibrating}, in agreement with the age from stellar models. %Alex: Given the errorbars, everything should be compatible !
%In an attempt to compare Gyrochronological ages with those from stellar models, we searched for rotational modulation on the star in the K2 lightcurve.
%We applied a Lomb-Scargle function \citep{press1989fast} to the lightcurve and detected significant long-duration periodicity that may be starspot rotation features. 
%However, due to differences in detrending methods, the PDC lightcurve \citep[Kepler Presearch Data Conditioning,][]{stumpe2012kepler} suggests variation on the order of 26d (corresponding to $\sim3.0 \pm 3$Gyr), hence we cannot confirm a specific rotation period. %Alex: So, is it worth mentioning?
A slow stellar rotation and old age are also supported by the upper limit of v$_{\rm rot}$ measured from HARPS spectra ($<2.7$~kms$^{-1}$, P$_{\rm min}>9.2$~d), and the non-detection of lithium ([Li/H]$<0.2$).

Recently, \cite{NissenYMg} showed that [Y/Mg] ratio can be used to estimate stellar ages.
This result was later confirmed by \cite{TucciMaia2016}, the age relation from which suggests an age of $8.1 \pm 2.8$ Gyr for K2-110.
Hence, all three methods point this star being an old field star, which may also explain its low metallicity.

%Ruth Angus gyro age: $A = \left( P / \left(B - V - 0.45 \right)^{0.31} \right) ^{1.82} $%0.40* \( B − V − 0.45 \)^{0.31}\)\)^{\(1/0.55\)}$
%For 5000K star (B-V=0.86), sub-25d periods from sub-3Gyr stars, so likely older than 3Gyr.
%50d LAM period consistent with extremely old star.
%The vsin i upper limit of $<2.7kms^{-1}$ is consistent with this slow rotation and therefore old age.

\subsection{Stellar composition}
The low metallicity of K2-110 is in agreement with the finding that small planets can form around metal-poor host stars \citep{buchhave2012abundance}.
However, the interior mass (16.5~M$_{\oplus}$) appears anomalously large for such a metal-poor star, with the median mass sitting above the mass-metallicity upper boundary as found by \cite{courcol2016upper} for Neptunian exoplanets.
%Alex: missing bib entry for this reference
Unusually, K2-110 does not show an enhancement in $\alpha$-elements (e.g. Si, Mg, etc) compared to other metal-poor planet hosts \citep{vardan2012} %Alex: missing bib entry for this reference
though it does show enhancement in the pure $\alpha$-element oxygen ([O/Fe]=0.35).

\subsection{TTVs and other planets}
Using a transit model generated from our \texttt{PASTIS} best fit, we searched for transit timing variations by iteratively shifting the flux model over each individual transit with a resolution of 2.6~s.
We detect no significant TTVs and are able to rule out their presence above an amplitude of 6 minutes to 3$\sigma$.

We also searched for potential other transiting planets in the system but found no significant signal. 
Injection \& retrieval tests enable us to rule out to $>90\%$ confidence the presence co-planar planets with orbits $<30$~d and radii $>1$~R$_{\oplus}$.
This, along with the location of K2-110 b at the metallicity-mass upper limit, suggests that the planet is likely solitary and contains the majority of K2-110's protoplanetary disc mass.

\subsection{Composition and formation}
\begin{figure}
\makebox[\columnwidth]{
 \includegraphics[width=\columnwidth]{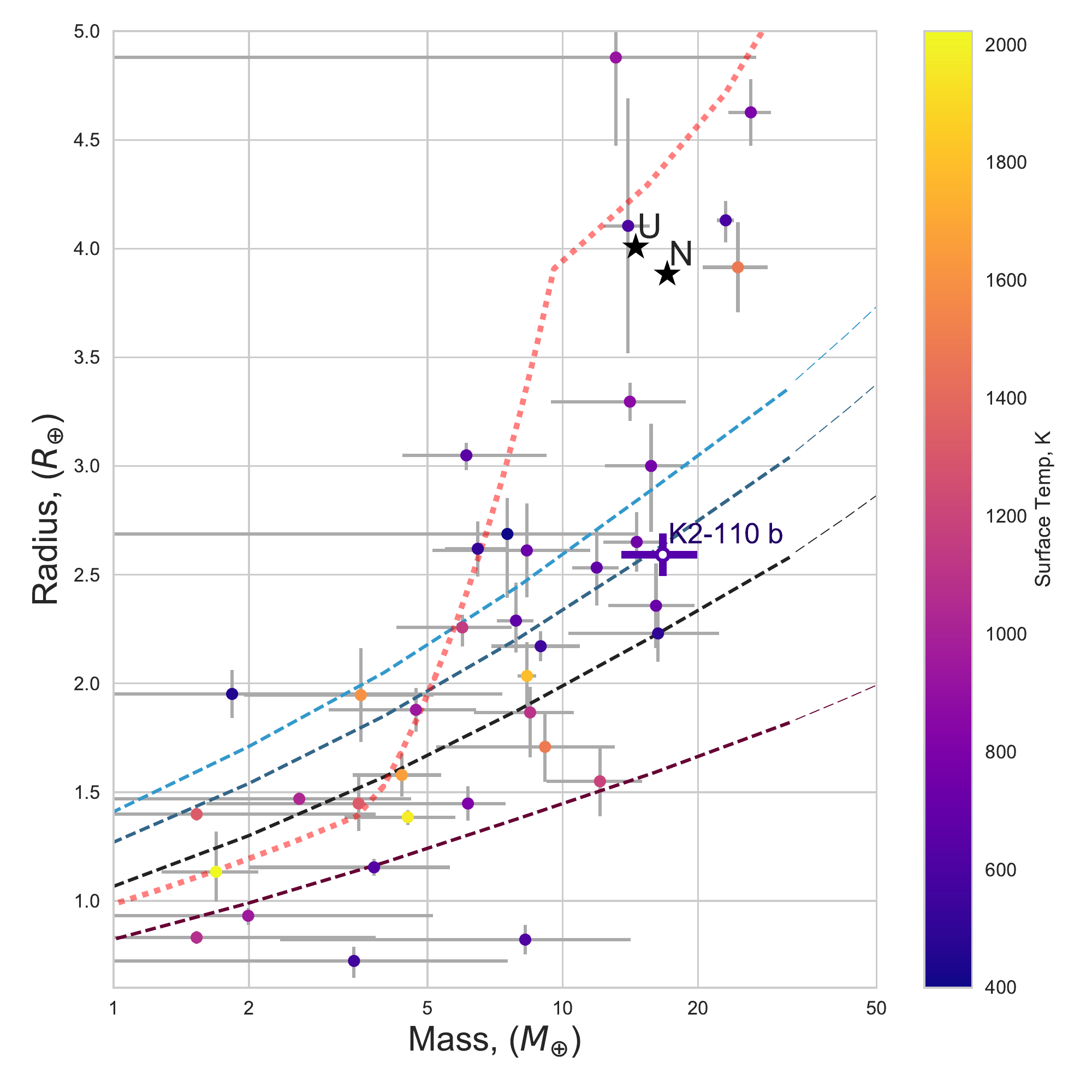}
}
\caption{K2-110 b (solid cross, right) compared to other super-Earth and Neptunian planets (data from \cite{exoplanetorg,marcy2014masses}). 
Mass-radius relations adapted from \cite{zengsass} for 100\%-Fe, Earth-like, 50\%-water and 100\% water  compositions (dashed lines from bottom to top).
A mass-radius relation for small exoplanets \citep{weiss2014mass} is also show (fine dashes).}
\label{MRdiag}
\end{figure}

With a mass of $16.7 \pm 3.2$~M$_{\oplus}$ and a radius of $2.6 \pm 0.1$~R$_{\oplus}$, this planet stands out as being one of the most massive exoplanets with a sub-Neptune radius (Fig. \ref{MRdiag}) detected so far.
Despite an Earth-like density of 5.2$\pm$1.2~g.cm$^{-3}$, a 2-layer iron-silicate composition model is unable to explain the density of K2-110 b.
Instead, either low-density volatiles such as water, an H-He atmosphere, or a combination of both must be present.
We explore the possibility of these degenerate compositions here.
Using the 3-layer solid exoplanet composition model of \cite{zengsass}\footnote{Accessed from https://www.cfa.harvard.edu/~lzeng/}, we compute that a $8.5$~M$_{\oplus}$ Earth-like interior of $\sim70\%$ MgSiO$_3$ and $\sim30\%$ Fe covered by 8~M$_{\oplus}$ of H$_2$O can explain the mass and radius of K2-110 b.
As well as surface molecular water, such a model would also require high-pressure water phases Ice VII, Ice X and superionic fluid \citep{zeng2014effect}.

In the alternate case, the models of \cite{adams2008ocean} and \cite{lopez2014understanding} both show that an earth-like $16.5$~M$_{\oplus}$ interior of iron core and silicate mantle can host a $\sim0.2$~M$_{\oplus}$, $\sim0.4$~R$_{\oplus}$ H-He atmosphere to produce the equivalent bulk density.
Intermediate compositions between these two boundary models are also possible.
This suggests the mass fraction of hydrogen is likely <1\% for K2-110 b.

%The accretion and subsequent de-gassing of a rocky exoplanet could produce as much as 6\% H\textsubscript{2} and 23\% H\textsubscript{2}O by mass, although much water could be retained by a silicate mantle. Hence, without the accretion

The growing population of high-density planets in the regime from 10 to 20~M$_{\oplus}$ also suggests that planets can exist in this region without accreting significant hydrogen.
This may therefore suggest that the minimum core accretion mass of $\sim10$~M$_{\oplus}$ \citep{pollack1996formation} is underestimated, or that processes exist to remove gaseous atmospheres post-accretion.
The unusual density of this planet also suggests that mass-radius relations \citep[e.g.][]{weiss2014mass}, should be used with extreme caution in the regime between terrestrial planets and gas giants (e.g. Fig. \ref{MRdiag}).

%For example, the mass and radius of EPIC-1166 could be acheived from a large $19M_/oplus$ core and a $<1M_/oplus$ outer layer of H/He (contributing 

%More massive than Neptune, but more than 3-times denser, despite significant heating (600C Tsurf).
%Consistent with 50\% MgSiO4 and 50\% H2O
%From Weng-Sasselov: $19.48R_e, 2.717M_e, Core radius fraction=0.6252 p_{cmb}/p_{centre}=0.2353$
%Or Fe $19.3, 2.73: CRF=0.61, CMF=0.73, p_{cmb}/p_{centre}=0.28$

%For 3-layer composition models (without H/He), it is likely 51-73\% H2O with <49\% silicate or <25\% iron, or a more likely composition of 60:30:10 H2O:Si:Fe
%How does adding 1\% H2 (0.19Me) increase the radius?

%This density and composition put EPIC-1166 on the boundary between high-density evaporated giant planet cores like Kepler-10c (http://arxiv.org/abs/1405.7881) and low-density Neptunes like HAT-26b (http://arxiv.org/abs/1010.1008).

%Same radius as GJ1214 (7Me) / Kepler-11c (3Me)/ Kepler-84 (30Me)

%H+He envelopes likely <1\% of planets' total masses (From radii only, http://iopscience.iop.org/article/10.1088/0004-637X/806/2/183/meta)

%Evaporation?
To investigate whether K2-110 b could have been formed with a substantial atmosphere that later evaporated, we looked at the potential mass loss from EUV.
Taking an average of the most EUV-luminous K-type star, we use the calculations of \citet{des2007diagram} and calculate a mass loss rate of $2\times10^9 ~$gs$^{-1}$ (Eq. 15), which corresponds to an upper limit of 0.13~M$_{\oplus}$ over 10~Gyr.
This is of similar magnitude to a maximum X-ray mass-loss rate computed from \cite{owen2012planetary}.
Therefore, using the Hydrogen atmosphere models mentioned previously, the host star is likely incapable of evaporating more than 0.2~R$_{\oplus}$ of Hydrogen from K2-110 b. 
Thus evaporation is unlikely to have contributed to the high density we see today.
Hence K2-110 b likely formed dense, potentially after the gas disc has been photoevaporated.

However, to form K2-110 b in situ at 0.1~AU would require a disc mass enhanced by 50 compared to solar values \citep{Schlichting2014}.
Hence, either material from more distant parts of the disc migrated inwards to build K2-110 b \citep[e.g.][]{chatterjee2013inside,hansen2012migration}, or the planet itself formed far out in the disc and migrated inwards \citep[e.g.][]{mcneil2010formation,kley2012planet}.

%The migration of rocky material into the inner solar system could produce a planet at this position. .
%Lee and Chaing paper - in situ formation of large neptunes - ar 0.1AU
%Can form planetesimals close-in during end of disc and only collect 1%
% Could produce 1% by outgassing
% Lindy elkins-tantum. ASU - outgassing of Water-rich core could produce 1%
% 
In the latter case, migration could have occurred through dynamic scattering, or through disc migration.
Better constraints on orbital eccentricity ($e=0.08\pm0.07$ in this analysis) could help point such an orbital migration mechanism, however we can rule out high eccentricities ($e>0.25$) typical of warm exoplanets caused by Kozai migration \citep{dawson2014class}.
On its relatively wide orbit of 14d, K2-110 b is also unlikely to have been to acted on by stellar tidal forces; a necessary component of dynamic migration \citep{fabrycky2007shrinking}.
Therefore, if K2-110 b migrated to its current position low-eccentricity disc migration is more likely.
However, the lack of a thick H atmosphere on K2-110 b is at odds with the migration of a $>10$~M$_{\oplus}$ planet through a gaseous disc.

On the other hand, models of the migration, impact and accretion of systems of compact planets or planetary embryos \citep{ida2008toward, boley2015situ} are able to explain both K2-110 b's orbit and its lack of significant hydrogen atmosphere \citep{liu2015giant}.
Improved orbital parameter measurements (e.g. misalignment and eccentricity) and statistical analyses of exoplanet populations could disentangle which scenario occurred.
%This hypothesis also supports the interpretation of an old age ($\sim8$Gyr) for this system, over which time the likelihood of dynamical instability is increased.

%EPIC-1166b is therefore a counter-example to the
%K-type stars have large EUV luminosities.
%dEeuv/dt = 1.35e22 (K-type, 14.7erg/cm2/s) or 4.3e34/Gyr
%$10^9-10^10g/s$
%dEeuv/dt = 4.25e21 (G-type, 4.6erg/cm2/s) or 1.34e34/Gyr
%Mass Loss for K-star: $~10^9 g/s$ (from http://www.aanda.org/articles/aa/pdf/2007/03/aa5014-06.pdf).
%Extrapolated over time  = 0.16Me over 10Gyr.
%600K surface T. Old, so likely long ation. K-star may be active. But less energy than most hot jupters.
%OTHER DISCUSSION POINTS:
%EPIC-1166's V-band magnitude of 11.9 and J, H \& K magnitudes of 10.19, 9.923 and 9.841 make it a candidate for re-observation. 

\section{Conclusion}
Using photometry from \textit{K2} we have detected a $2.6\pm0.1$~R$_{\oplus}$ planet orbiting an early K-dwarf.
RV observations with HARPS and HARPS-N have confirmed K2-110 b (EPIC 212521166) as a planet and measured its mass to be $16.7\pm3.2$~M$_{\oplus}$.
The corresponding bulk density suggests K2-110 b has a large rocky interior and is hydrogen-poor, with <1\% of its mass in a hydrogen atmosphere. 
Alternatively, the planet could be volatile-rich, with up to 9~M$_{\oplus}$ of H$_{2}$O.
Our analysis means K2-110 b is now one of the best-characterised sub-Neptune planets with a radius and mass constrained to 4\% and 20\% respectively. Future observations will improve our understanding of the bulk composition and migration of this planet.

%k_b = 1.38064852e23; Teq=1400 ; Rp=2.6*6400000 ; mu = 2.2 ; Mp = 16.7*5.96e24 ; g = (6.67e-11*Mp)/(Rp**2) ; R_2 = 0.713*695500000 ; t_14 = 3.22*3600 ; Hmag= 9.923 ; F = np.power(2.512,0.0-1*Hmag)/1024.0
%SN_atmos_1166 = (Rp*(k_b*Teq*(1.0/mu)*(1.0/g)*np.sqrt(F*t14))/(R_s**2)
%Teq=500 ; Rp=2.85*6400000 ; mu = 2.2 ; Mp = 16.7*5.96e24 ; g = (6.67e-11*Mp)/(Rp**2) ; R_2 = 0.828*695500000 ; t_14 = 4.81*3600 ; Hmag= 7.203 ; F = np.power(2.512,0.0-1*Hmag)/1024.0
%SN_atmos_HD3167c =  (Rp*(k_b*Teq*(1.0/mu)*(1.0/g)*np.sqrt(F*t14))/(R_s**2)

\begin{acknowledgements}
We are grateful to the pool of HARPS observers who conducted part of the visitor-mode observations at La Silla Observatory: Fatemeh Motalebi, Aurélien Wyttenbach, Baptiste LaVie, Pedro Figueira, Alessandro Sozzetti, and Ana\"el Wunsche.
%HPO acknowledges enlightening discussions with Matt Mutter \& Gavin Coleman, and funding from a Warwick University Chancellors Scholarship.
The Portuguese team acknowledges the support from the Fundação para a Ciência e Tecnologia (FCT) through national funds and by FEDER through COMPETE2020 by grants UID/FIS/04434/2013 \& POCI-01-0145-FEDER-007672, PTDC/FIS-AST/1526/2014 \& POCI-01-0145-FEDER-016886 and PTDC/FIS-AST/7073/2014\& POCI-01-0145-FEDER-016880. S.C.C.B., N.C.S. e S.G.S. also acknowledge support from FCT through Investigador FCT contracts numbers IF/01312/2014/CP1215/CT0004, IF/00169/2012/CP0150/CT0002 and IF/00028/2014/CP1215/CT0002. V.Zh.A. and E.D.M. also acknowledge support from FCT through Investigador FCT contracts IF/00650/2015/CP1273/CT0001, IF/00849/2015/CP1273/CT0003, and by the fellowship SFRH/BPD/70574/2010, SFRH/BPD/76606/2011 funded by FCT (Portugal) and POPH/FSE (EC). 
S.H. acknowledges support by the fellowship PD/BD/128119/2016 funded by FCT (Portugal) and POPH/FSE (EC). A.S. was supported by the EU under a Marie Curie Intra-European Fellowship for Career Developmentwith reference FP7-PEOPLE-2013-IEF, number 627202.
AS is supported by the EU under a Marie Curie Intra-European Fellowship for Career Development with reference FP7-PEOPLE-2013-IEF, number 627202.
DJAB acknowledges support from the UKSA and the University of Warwick. 
J.L-B  acknowledges support from the Marie Curie Actions of the European Commission (FP7-COFUND)
JMA acknowledges funding from the European Research Council under the ERC Grant Agreement n. 337591-ExTrA.
D.J.A. and D.P acknowledge funding from the European Union Seventh Framework programme (FP7/2007- 2013) under grant agreement No. 313014 (ETAEARTH). 
OD acknowledges support by CNES through contract 567133 P.A.W acknowledges the support of the French Agence Nationale de la Recherche (ANR), under programme ANR-12-BS05-0012 Exo-Atmos".

Some of the data presented in this paper were obtained from the Mikulski Archive for Space Telescopes (MAST). STScI is operated by the Association of Universities for Research in Astronomy, Inc., under NASA contract NAS5-26555. Support for MAST is provided by NASA grant NNX09AF08G \& by other grants and contracts.
This paper includes data collected by the \textit{Kepler} mission. Funding for the \textit{Kepler} mission is provided by the NASA Science Mission directorate.
\end{acknowledgements}

\bibliographystyle{aa}
\bibliography{Refs} 

\section{Appendices}

\onecolumn

\onecolumn

\begin{landscape}
\begin{table}
\centering
\caption{Raw radial velocity data from HARPS and HARPS-N.}     
\label{RVtab}
\begin{tabular}{c c c c c c c c c c c c c}  
\hline
Time  &  RV & $\sigma$ RV & FWHM  & $\sigma$ FWHM  & BIS & $\sigma$ BIS & \logrhk & $\sigma$ \logrhk  & S/N 50 & Texp & Instrument & $\Delta$RV$_{\rm moon}$\\
%\hline

[UTC]  &  [\kms] & [\kms]  & [\kms]  & [\kms]  & [\kms] & [\kms] & [dex] & [dex] & ~ & [s] & ~ \\
\hline
57451.737583  &  -21.6236  &  0.0012  &  5.940  & 0.002  &  0.019  &  0.002  &  -4.940  &  0.012  &  68.1  &  3600  &  HARPS & -\\
57457.817274  &  -21.6338  &  0.0018  &  5.936  & 0.004  &  0.006  &  0.003  &  -5.027  &  0.030  &  45.9  &  3600  &  HARPS & -\\
57458.702389  &  -21.6409  &  0.0015  &  5.942  & 0.003  &  0.012  &  0.002  &  -4.994  &  0.020  &  56.1  &  3600  &  HARPS & -\\
57458.869113  &  -21.6442  &  0.0016  &  5.937  & 0.003  &  0.018  &  0.002  &  -5.029  &  0.026  &  53.8  &  3600  &  HARPS & -\\
57459.739145  &  -21.6362  &  0.0027  &  5.930  & 0.005  &  0.016  &  0.004  &  -4.985  &  0.048  &  33.5  &  3600  &  HARPS & -\\
57460.725595  &  -21.6375  &  0.0017  &  5.946  & 0.003  &  0.023  &  0.003  &  -4.980  &  0.025  &  48.5  &  3600  &  HARPS & -\\
57460.826006  &  -21.6342  &  0.0015  &  5.943  & 0.003  &  0.026  &  0.002  &  -4.993  &  0.021  &  54.9  &  3600  &  HARPS & -\\
57461.855999  &  -21.6381  &  0.0013  &  5.943  & 0.003  &  0.015  &  0.002  &  -5.034  &  0.024  &  65.2  &  3600  &  HARPS & -\\
57462.825026  &  -21.6326  &  0.0012  &  5.944  & 0.002  &  0.022  &  0.002  &  -4.995  &  0.019  &  70.3  &  3600  &  HARPS & -\\
57463.862782  &  -21.6274  &  0.0016  &  5.939  & 0.003  &  0.018  &  0.002  &  -5.006  &  0.027  &  54.2  &  3600  &  HARPS & -\\
57464.704531  &  -21.6331  &  0.0024  &  5.935  & 0.005  &  0.014  &  0.004  &  -5.103  &  0.048  &  36.9  &  3600  &  HARPS & -\\
57464.793876  &  -21.6361  &  0.0030  &  5.947  & 0.006  &  0.024  &  0.004  &  -5.132  &  0.070  &  30.3  &  3600  &  HARPS & -\\
57465.710199  &  -21.6277  &  0.0019  &  5.946  & 0.004  &  0.023  &  0.003  &  -5.017  &  0.028  &  44.0  &  3600  &  HARPS & -\\
57465.823631  &  -21.6273  &  0.0022  &  5.947  & 0.004  &  0.007  &  0.003  &  -4.978  &  0.033  &  38.0  &  3600  &  HARPS & -\\
57478.825107  &  -21.6258  &  0.0016  &  5.936  & 0.003  &  0.014  &  0.002  &  -5.037  &  0.028  &  52.7  &  3600  &  HARPS & -\\
57567.639306  &  -21.6358  &  0.0019  &  5.984  & 0.004  &  0.019  &  0.003  &  -5.094  &  0.045  &  48.3  &  3600  &  HARPS & -\\
57607.484484  &  -21.6299  &  0.0021  &  5.948  & 0.004  &  0.022  &  0.003  &  -4.959  &  0.033  &  42.4  &  3600  &  HARPS & -\\
\hline
57521.558771  &  -21.6353  &  0.0027  &  5.867  &  0.004  &0.015  & 0.005  &    -4.856  &  0.035  &  36.6  &  1800  &  HARPS-N & 0.00018\\
57525.560510  &  -21.6441  &  0.0037  &  5.849  &  0.006  &0.021  & 0.007  &    -4.921  &  0.061  &  28.2  &  1800  &  HARPS-N & 0.00299\\
57526.564378  &  -21.6413  &  0.0035  &  5.841  &  0.005  & 0.022  & 0.007  &   -4.972  &  0.063  &  29.2  &  1800  &  HARPS-N & 0.00205\\
57528.543419  &  -21.6371  &  0.0032  &  5.908  &  0.005  & 0.082  & 0.006  &   -5.000  &  0.054  &  32.9  &  1800  &  HARPS-N & 0.02834\\
57529.513743  &  -21.6415  &  0.0025  &  5.879  &  0.004  & 0.042  & 0.005  &   -4.945  &  0.037  &  38.5  &  1800  &  HARPS-N & 0.01055\\
57557.437764  &  -21.6305  &  0.0056  & 5.899  &  0.008  & -0.019  & 0.011  &    -5.007  &  0.113  &  20.5  &  1800  &  HARPS-N & -0.03654\\
57559.442078  &  -21.6457  &  0.0043  &  5.903  &  0.006  &   0.006  & 0.009  & -4.862  &  0.061  &  25.8  &  1800  &  HARPS-N & -0.03128\\
57562.439809  &  -21.6297  &  0.0027  &  5.867  &  0.004  &   0.005  & 0.005  & -4.943  &  0.041  &  37.1  &  1800  &  HARPS-N & -0.00407\\
57563.419297  &  -21.6363  &  0.0041  &  5.860  &  0.006  &   0.013  & 0.008  & -4.882  &  0.063  &  26.2  &  1800  &  HARPS-N & -0.00142\\
57564.413990  &  -21.6313  &  0.0088  &  5.836  &  0.013  &   0.042  & 0.018  & -4.802  &  0.120  &  15.3  &  1800  &  HARPS-N & 0.01507\\
57565.442256  &  -21.6316  &  0.0032  &  5.857  &  0.005  &   0.005  & 0.006  & -5.001  &  0.060  &  32.1  &  1800  &  HARPS-N & 0.00090\\
\hline
\end{tabular}
\end{table}
\end{landscape}

\begin{table}
\caption{Stellar abundances}     
\label{abunds}
\begin{tabular}{c l l c  c c l l}     % 3 columns 
\hline      
% To combine 4 columns into a single one 
Elem & [X/H] & error & ~ & Elem & [X/H] & error\\ 
\hline
O & $0.009$ & $0.080$ & ~ & Cu & $-0.305$ & $0.048$ \\ 
\ion{Na}{i}  &  -0.334  &  0.046 & ~ & \ion{Zn}{ii} & $-0.323$ & $0.101$ \\ 
\ion{Mg}{i}  &  -0.274  &  0.060 & ~ & Sr & $-0.338$ & $0.050$ \\ 
\ion{Al}{i}  &  -0.157  &  0.035 & ~ & \ion{Y}{ii} & -0.425 & 0.093 \\
\ion{Si}{i}  &  -0.260  &  0.039 & ~ & Zr & $-0.290$ & $0.080$ \\ 
\ion{Ca}{i}  &  -0.223  &  0.062 & ~ & \ion{Ba}{ii} & $-0.452$ & $0.010$ \\ 
\ion{Sc}{ii}  &  -0.276  &  0.064 & ~ & \ion{Ce}{i} & $-0.131$ & $0.327$ \\ 
Ti  &  -0.205  &  0.092 & ~ & Mg/Si & 1.191 & --\\
%\ion{V}{i}   &  0.100  &  0.112\\
Cr  &  -0.279  &  0.074 & ~ & O/Fe & 0.35 & 0.08\\
\ion{Mn}{i}  &  -0.331  &  0.077 & ~ & A(Li) & $<0.2$ & -- \\
\ion{Co}{i}  &  -0.272  &  0.031 & ~ & -- & ~ & ~ \\
% \ion{Ni}{i}  &  -0.336  &  0.023 
% %C & $0.358$ & $0.050$ \\ 
% Cu & $-0.305$ & $0.048$ \\ 
% \ion{Zn}{ii} & $-0.323$ & $0.101$ \\ 
% Sr & $-0.338$ & $0.050$ \\ 
% %\ion{Y}{ii} & $-0.433$ & $0.155$ \\
% \ion{Y}{ii} & -0.425 & 0.093 \\
% Zr & $-0.290$ & $0.080$ \\ 
% \ion{Ba}{ii} & $-0.452$ & $0.010$ \\ 
% \ion{Ce}{i} & $-0.131$ & $0.327$ \\ 
% Mg/Si & 1.191 & --\\
% O/Fe & 0.35 & 0.08\\
% A(Li) & $<0.2$ & -- \\
\hline
%From Vardan. Similar Mg/Si to HD 97658 b (Elisa Delgado Mena)
\end{tabular}
\end{table}
%The star apparently is not alpha enhanced. [alpha/Fe] ~ 0.1 dex. It is strange, but Vanadium and Ti (not TiII) show some enhancement. The enhancement in VI is quite large, although the error for this element is also the largest one. In our previous works  we found that [V/Fe] ratio increases with the decrease of Teff (for cool stars), but it was due to 'incorrect' Teffs. I was expecting with Maria's line-list to solve this problem, but maybe not yet. ask Vardan about that, he is the expert of that.

\begin{table*}
\caption{List of free parameters used in the \texttt{PASTIS} analysis of the light curves, radial velocities and SED with their associated prior and posterior distribution. }
\begin{center}
\begin{tabular}{lcc}
\hline
\hline
Parameter & Prior & Posterior\\
\hline
\multicolumn{3}{l}{\it Orbital parameters}\\
 &&\\
Orbital period $P$ [d] & $\mathcal{U}(13.8;13.9)$ & 13.86375 $\pm$ 2.6$\times10^{-4}$\\
Epoch of first transit T$_{0}$ [BJD$_{\rm TDB}$] - $2.45\times10^6$ & $\mathcal{U}(7275.2 ; 7275.5)$ & 7275.32991 $\pm$ 6.3$\times10^{-4}$\\
Orbital eccentricity $e$ & $\beta(0.867;3.03)$ & 0.091 $\pm$  0.089\\
Argument of periastron $\omega$ [\degr] & $\mathcal{U}(0;360)$ & 90$^{_{+180}}_{^{-64}}$ \\
Inclination $i$ [\degr] & $\mathcal{S}(70;90)$ & 89.35$^{_{+0.41}}_{^{-0.24}}$\\
 &&\\
\hline
\multicolumn{3}{l}{\it Planetary parameters}\\
 &&\\
Radial velocity amplitude $K$ [\ms] & $\mathcal{U}(0;1000)$ & 5.5 $\pm$ 1.1\\
Planet-to-star radius ratio $k_{r}$ & $\mathcal{U}(0; 1)$ & 0.0333$\pm6.6\times10^{-4}$\\
 &&\\
\hline
\multicolumn{3}{l}{\it Stellar parameters}\\
 &&\\
Effective temperature \teff\ [K] & $\mathcal{N}(4960:60)$ & 5010 $\pm$ 50\\
Surface gravity \logg\ [g.cm$^{-2}$] & $\mathcal{N}(4.58;0.13)$ & 4.60 $\pm$ 0.03\\
Iron abundance \met\ [dex] & $\mathcal{N}(-0.34 ; 0.03)$ & -0.34 $\pm$ 0.03\\
Reddening E(B-V) [mag] & $\mathcal{U}(0;1)$ & 0.019$^{_{+0.019}}_{^{-0.013}}$\\
Systemic radial velocity $\upsilon_{0}$ [\kms] & $\mathcal{U}(-25, -15)$ & -21.6331 $\pm 9\times10^{-4}$\\
Distance to Earth $d$ [pc] & $\mathcal{P}(2;10;1000)$ & 118.0 $\pm$ 3.6\\
 &&\\
\hline
\multicolumn{3}{l}{\it Instrumental parameters}\\
 &&\\
HARPS radial velocity jitter [\ms] & $\mathcal{U}(0;1000)$ & 3.1 $\pm$ 1.0\\
HARPS-N radial velocity jitter [\ms] & $\mathcal{U}(0;1000)$ & 3.2$\pm$2.3\\
HARPS -- HARPS-N radial velocity offset [\ms] & $\mathcal{U}(-100;100)$ & -4.2 $\pm$ 1.8\\
SED jitter [mag] & $\mathcal{U}(0;1)$ & 0.021 $\pm$ 0.019\\
\textit{K2} jitter [ppm] & $\mathcal{U}(0;10000)$ & 41.2 $\pm$ 4.6\\
\textit{K2} contamination [ppt] & $\mathcal{N}_{\mathcal{U}}(0;5;0;100)$ & 3.4$^{_{+3.6}}_{^{-2.4}}$\\
%\textit{K2} flux normalisation & $\mathcal{U}(0.999;1.001)$ &  $\pm$ 0.025\\
\hline
\hline
\end{tabular}
\tablefoot{$\mathcal{N}(\mu;\sigma^{2})$ is a normal distribution with mean $\mu$ and width $\sigma^{2}$, $\mathcal{U}(a;b)$ is a uniform distribution between $a$ and $b$, $\mathcal{N}_{\mathcal{U}}(\mu;\sigma^{2},a,b)$ is a normal distribution with mean $\mu$ and width $\sigma^{2}$ multiplied with a uniform distribution between $a$ and $b$, $\mathcal{S}(a,b)$ is a sine distribution between $a$ and $b$, $\beta(a;b)$ is a Beta distribution with parameters $a$ and $b$, and $\mathcal{P}(n;a;b)$ is a power-law distribution of exponent $n$ between $a$ and $b$.}
\tablebib{The choice of prior for the orbital eccentricity is described in \citet{kipping2013parametrizing}.} %citation to Kipping (2013)
\end{center}
\label{PASTISparams}
\end{table*}%

\end{document}